\documentclass[%
 amsmath,amssymb,
 aps,
prb,
floatfix,
twocolumn
]{revtex4-1}

\usepackage{CJKutf8}

\usepackage{graphicx}
\usepackage{color}
\usepackage{dcolumn}
\usepackage{bm}
\usepackage{wasysym}
\usepackage{bbold}
\usepackage{hyperref}

\begin{document}

\title{Symmetry, distorted bandstructure, and spin-orbit coupling\\ of (group-III) metal-monochalcogenide monolayers} 

\author{Pengke Li (\begin{CJK*}{UTF8}{gbsn}李鹏科\end{CJK*})}
\email{pengke@umd.edu}
\author{Ian Appelbaum}
\email{appelbaum@physics.umd.edu}
\affiliation{Department of Physics and Center for Nanophysics and Advanced Materials, University of Maryland, College Park, MD 20742}

\begin{abstract}
The electronic structure of (group-III) metal-monochalcogenide monolayers exhibits many unusual features. Some, such as the unusually distorted upper valence band dispersion we describe as a `caldera', are primarily the result of purely orbital interactions. Others, including spin splitting and wavefunction spin-mixing, are directly driven by spin-orbit coupling. We employ elementary group theory to explain the origins of these properties, and use a tight-binding model to calculate the phenomena enabled by them, such as band-edge carrier effective g-factors, optical absorption spectrum, conduction electron spin orientation, and a relaxation-induced upper valence band population inversion and spin polarization mechanism.\vspace{-12pt}
\end{abstract}

\pacs{Valid PACS appear here}

\maketitle

\section{\label{sec:intro} Introduction}
\vspace{-12pt}
Three-dimensional van~der~Waals solids from the (group-III) metal monochalcogenide layered semiconductor family $MX$ (where $M$ is Ga or In, and $X$ is S, Se, or Te) have been intensively investigated by both experiment and theory for many decades. The classical literature on this subject\footnote{A pre-1994 bibliography of more than 600 papers on GaSe can be found at \href{http://www.dtic.mil/cgi-bin/GetTRDoc?AD=ADA279064}{http://www.dtic.mil/cgi-bin/GetTRDoc?AD=ADA279064}} contains reports of measurements on optical absorption, photo- and electroluminescence, photoconductivity, radiative recombination, electrical conductivity, and Hall effect \cite{Bube_PR59,Fielding_JPCS59,Fischer_JPCS62,Springford_PPS63,Brebner_JPCS64, Leung_JPCS65,Akhundov_PSSB66,Mercier_JL73}. Even conduction electron spin polarization via optical orientation was carried out to study spin-dependent carrier dynamics.\cite{Ivchenko_JETP77, *Ivchenko_ZhETF77, Gamarts_JETP77, *Gamarts_ZhETF77} On the theory side, we find a first attempt to derive the band structure based on symmetry analysis and a simple few-band model occurring over fifty years ago.\cite{Fischer_Helvetica63} Since then, many detailed bandstructure calculations have appeared, utilizing the empirical pseudopotential method \cite{Schluter_Nuovo71, Schluter_PRB76, Depeursinge_SSC78} or tight binding formalism.\cite{Bassani_INC67, Kamimura_JPSJ68, McCanny_JPC77, Balzarotti_SSC77, Nagel_JPC79, Doni_INC79, Robertson_JPC79, Camara_PRB02}

However, it was not until the recent search for beyond-graphene\cite{Geim_NatMat07, Tan_PRL07} two-dimensional semiconductors (such as transition metal dichalcogenides\cite{Splendiani_NanoLett10,Mak_PRL10,Song_PRL13,Xu_NatPhys14} and phosphorene\cite{Liu_ACSNano2014, Li_NatureNano2014, Li_PRB14, Wang_NatureNano2015}) that this class of material was experimentally realized down to few- or mono-layer thickness by mechanical exfoliation from bulk crystals. Experimentally, GaS and GaSe ultrathin layer transistors have been demonstrated.\cite{Late_AdvMat12} Photoluminescence measurements show exciton features and reduction of optical efficiency when sample thickness decreases. \cite{Pozo_2D15,Schwarz_Nano14} Circularly polarized photoluminescence reveals spin dynamics in nanoslabs.\cite{Tang_JAP15, Tang_PRB15}  In some cases, monolayer $MX$ can even be synthesized epitaxially on silicon\cite{Vinh_JAP97} or non-epitaxially on insulating substrates such as SiO$_2$ via vapor phase deposition, with quality that rivals exfoliated material. \cite{Lei_NanoLett13,Li_SciRep14}

Despite this recent explosion of experimental results with monolayer $MX$, the theoretical establishment has relied almost exclusively on sophisticated \textit{ab initio} methods to model electronic structure. In particular, an unusual distortion of the highest valence band (sometimes called an `upside-down Mexican hat'\cite{Zolyomi_PRB13} or `sombrero') is predicted to create an indirect bandgap and a density-of-states singularity at or near the band edge.\cite{Wickramaratne_JAP15} This feature, which is more appropriately called a `caldera', vanishes in the bulk.\cite{Rybkovskiy_PRB14} First-principles schemes have made other fascinating predictions, such as spontaneous magnetism in p-type monolayer GaSe.\cite{Cao_PRL15, Wu_arxiv14} However, the underlying fundamental physics at the root of many intriguing properties is obscured by these brute-force numerical approaches, leaving many elementary questions without satisfactory answers.

Our aim in the present work is to investigate the underlying symmetries responsible for the many unique properties that are common to all $MX$ monochalcogenide monolayers sharing the same type of lattice structure and zone-center band-edge states. Using elementary group theory, we reveal the origin of the extraordinary `caldera' shape valence band edge, and examine various important phenomena made possible by the spin-orbit symmetry. These include orbital degeneracy breaking in the valence band, $k$-cubic dependence of the lowest order Dresselhaus splitting, orbital magnetism and effective Land\'{e} \textit{g}-factor, and spin dynamics during optical orientation. To assist the reader in acquiring an intuitive and quantitative understanding of our theory, we take the case of GaSe as a specific example in numerical calculations, using an empirical tight-binding model following Refs.~[\onlinecite{Robertson_JPC79}] and [\onlinecite{Camara_PRB02}] and incorporating on-site spin-orbit coupling parameters,\cite{Chadi_PRB77} as well as \textit{ab initio} density-functional theory (DFT) with the {\sc Quantum ESPRESSO} package.\cite{QE-2009} We emphasize that these numerical procedures are implemented only as a verification of the underlying physics determined by symmetry, which remains robust regardless of numerical details (such as the choice of functional or pseudopotential in DFT).

This paper is organized as follows: We start with background information on the essential group theory in Sec. \ref{sec:sym}, and with the assistance of the nearly free electron model, we analyze the symmetry of the spin-independent bandstructure, revealing the interactions causing an unusual valence band distortion: the `caldera'. In Sec. \ref{sec:symsoi}, we include spin-orbit interaction and investigate its effect on broken band degeneracy, eigenfunction composition, and spin splitting. In Sec. \ref{sec:gfactor} we show how orbital diamagnetism and valley-spin coupling result from the spin-mixed wavefunction symmetry. Finally, in Sec. \ref{sec:opt_ori} we discuss conduction electron spin orientation by optical excitation, including relaxation dynamics causing spin-polarized population inversion in the lower valence band and three-level spin pumping of the upper valence band.

\section{\label{sec:sym} Symmetry \textit{sans} spin}

\begin{figure}
\includegraphics[width=3.2in]{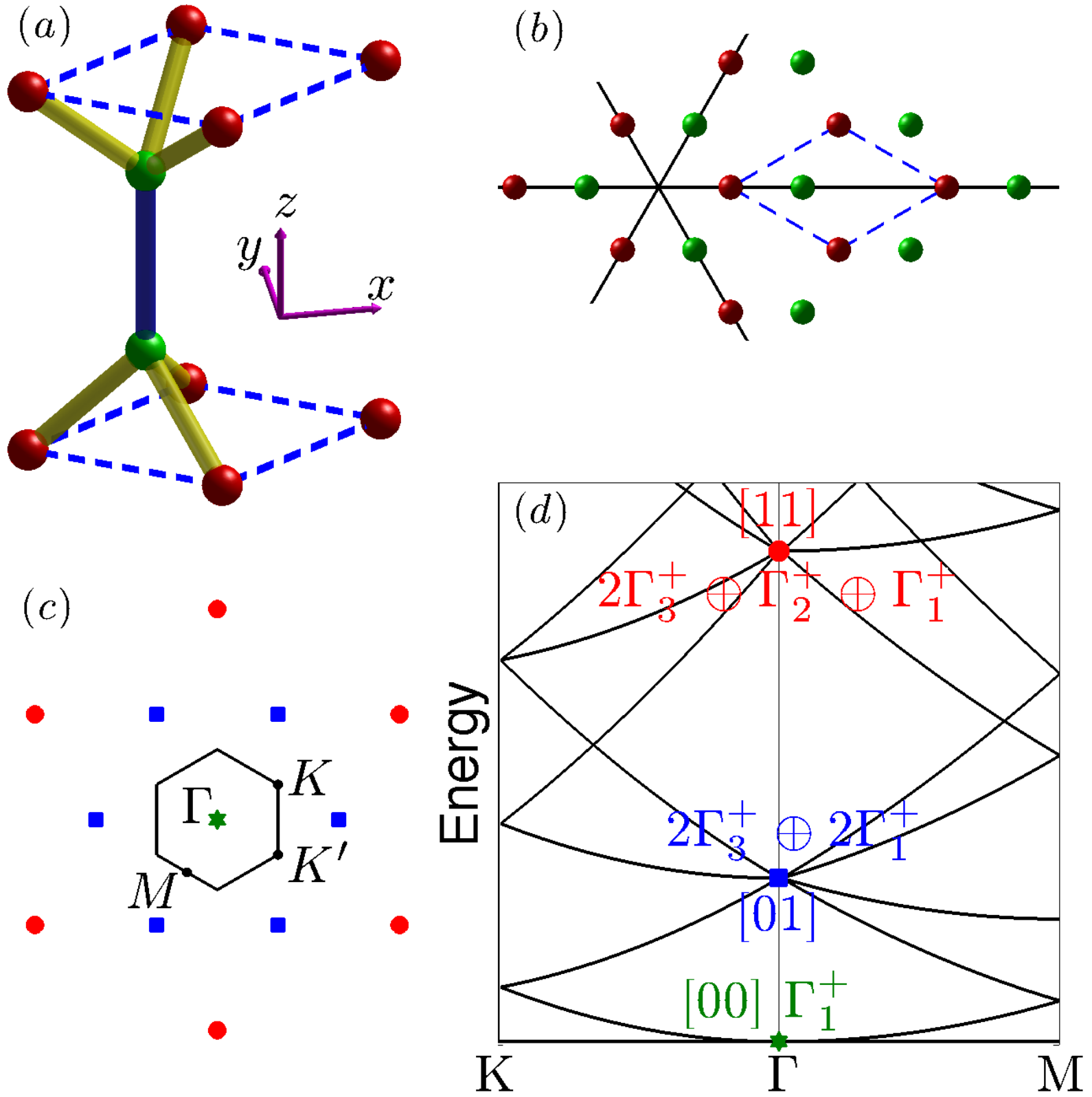}
\caption{Unit cell of single-layer group-III metal monochalcogenide in (a) perspective- and (b) plan-view. 
The red and green spheres correspond to chalcogen anions  and metal cations, respectively.  The dashed blue frame represents the unit cell boundaries. 
The three solid lines in (b) are axes for $180^\circ$ in-plane rotation operations. 
Panel (c) shows reciprocal lattice points and reduced Brillouin zone, where different colored markers indicate symmetry-related points of different zone-center character, resulting in degeneracies of $\Gamma$-point planewave eigenstates in the (d) nearly-free electron bandstructure along $K-\Gamma-M$ axes. \label{fig:lattice}}
\end{figure}

The unit cell of the monolayer metal-monochalcogenide hexagonal lattice is composed of two group-III metal and two chalcogen atoms, forming an upper and lower sublayer related by in-plane mirror reflection symmetry [Fig. \ref{fig:lattice}(a)]. Within each layer, the two types of atoms are covalently bonded and arranged alternately at the honeycomb lattice sites [Fig. \ref{fig:lattice}(b)]. The honeycomb lattice is buckled such that metal atoms are closer to the opposite sublayer. In this configuration, the two sublayers are tightly bound by adjacent metal atoms [blue bond in Fig. \ref{fig:lattice}(a)]. 

The first Brillouin zone is shown by the hexagon inscribed within the reciprocal lattice in Fig. \ref{fig:lattice}(c), together with high symmetry points $\Gamma$, $K$ ($K^\prime$) and $M$. Reciprocal lattice points equidistant from the origin can be divided into three sets according to their symmetry, denoted in Fig. \ref{fig:lattice}(c) by three different types of markers (green hexagram, blue square and red circle), corresponding to the three types of zone-center states in the nearly free electron band structure shown in Fig. \ref{fig:lattice}(d). Before proceeding with a detailed discussion of electronic properties, we first present a brief symmetry analysis of the system using group theory.

\subsection{\label{sec:group} Group theory }

The space group of single-layer $MX$ is symmorphic, with its point group $D_{3h}$ including twelve symmetry operators divided into six classes, denoted by $\mathbb{C}_{1\sim6}$ in Table~\ref{tab:CharacterTable}. The identity operator $E$ is in class $\mathbb{C}_1$. $\mathbb{C}_2$ includes the $\pm120^\circ$ rotations along an out-of-plane axis [through the position of any atom or the honeycomb center in Fig. \ref{fig:lattice}(b)]. $\mathbb{C}_3$ is composed of a $180^\circ$ rotation along the three axes [solid lines in Fig. \ref{fig:lattice}(b)] within the plane bisecting the two sublayers. $\mathbb{C}_4$ includes the in-plane mirror reflection $\sigma_h$. $\mathbb{C}_5$ and $\mathbb{C}_6$ take into account operators from the product of $\sigma_h$ and those in $\mathbb{C}_2$ and $\mathbb{C}_3$, respectively. 

\begin{table} [h!]
\caption{Character table of the $\Gamma$-point $D_{3h}$ group.  
$\Gamma_{1-6}$ in parenthesis are corresponding IRs in the convention of Bethe notation, in which the extra IRs after the double group extension are $\Gamma_{7-9}$ not listed here. The basis functions, $s$ and $p_{x,y,z}$ orbital configurations, and important invariants are also listed. The assignment of plus and minus superscripts to representations follows the convention of even and odd parity with respect to the operation of in-plane mirror reflection $\sigma_h$ of $\mathbb{C}_4$. Superscripts on orbitals indicate relative sign between orbitals on different (upper/lower) sublayers. Note that $\mathbf{k} =(k_x,k_y)= k(\cos\phi, \sin\phi) $}
\label{tab:CharacterTable}
\renewcommand{\arraystretch}{1.1}
\begin{tabular}{c|cccccc|c|c|c}
\hline \hline
&$\mathbb{C}_1$&$\mathbb{C}_2$&$\mathbb{C}_3$&$\mathbb{C}_4$&$\mathbb{C}_5$&$\mathbb{C}_6$ & basis & orbitals & invariants\\ \hline
$\Gamma_1^+(\Gamma_1)$  & 1 &1& 1& 1 &1 &1 & $\mathbb{1}$ &$s^+$, $p_z^-$ & $k^2$ \\
$\Gamma_2^+(\Gamma_2)$  & 1 &1&-1 &1 &1 &-1 & $xy$ & & $\sigma_z$, $\sin 3\phi$\\
$\Gamma_3^+(\Gamma_6)$  & 2 &-1& 0& 2 &-1 &0  & \{$x$,$y$\} & \{$p_x^+$,$p_y^+$\} & \{$k_x$,$k_y$\} \\\hline
$\Gamma_1^-(\Gamma_4)$  & 1 &1& 1& -1 &-1 &-1 & $xyz$ & \\
$\Gamma_2^-(\Gamma_3)$  & 1 &1&-1 &-1 &-1 &1 & $z$ & $s^-$, $p_z^+$\\
$\Gamma_3^-(\Gamma_5)$  & 2 &-1& 0& -2 &1 &0 & \{$xz$,$yz$\} & \{$p_x^-$,$p_y^-$\} & \{$\sigma_x$,$\sigma_y$\}\\
\hline
\end{tabular}
\end{table}

The Brillouin zone center $\Gamma$-point has the same symmetry of the point group $D_{3h}$. The six irreducible representations (IRs), denoted by $\Gamma_{1,2,3}^\pm$, have the characters given in Table~\ref{tab:CharacterTable}. The plus (minus) superscript reflects the even (odd) parity with respect to $\sigma_h$. Corresponding to each IR, we give in Table~\ref{tab:CharacterTable} the lowest order basis functions which we will use to describe the $\Gamma$-point states based on their wavefunction symmetries.

For intuitive understanding, and to aid numerical calculation of the band structure in the tight-binding formalism, we also provide in Table~\ref{tab:CharacterTable} the symmetries of $s$ and $p$ atomic orbital combinations, following Ref.~[\onlinecite{Robertson_JPC79}]. Here, similar to the IR labeling convention discussed above, even (odd) parity of orbital configuration within the unit cell upon application of $\sigma_h$ is denoted by the plus (minus) superscript. For example, $p_z^-$ of $\Gamma_1^+$ consists of $p_z$ orbitals of atoms in the two sublayers with opposite wavefunction phase orientation. Because the atoms are from group-III and VI, these $s$ and $p$ orbitals dominate the low energy electronic structure covering the critical band gap region. Their limited combinations in Table~\ref{tab:CharacterTable} indicate that only four IRs ($\Gamma_1^+$, $\Gamma_2^-$, $\Gamma_3^+$ and $\Gamma_3^-$) are relevant through out this paper. 

The right-most column of Table~\ref{tab:CharacterTable} lists several low order invariants according to their behavior under the twelve symmetry operations. For example, the function $\sin3\phi$ belongs to $\Gamma_2^+$, where $\phi$ is the polar angle of the 2D vector $\mathbf{k}$ with respect to the $\Gamma-M$ axis. It transforms to $-\sin3\phi$ under the six operations in classes $\mathbb{C}_3$ and $\mathbb{C}_6$ that involve rotations along in-plane axes. These invariants are low-order terms in the expansion of the Hamiltonian based on $\mathbf{k}\cdot\mathbf{\hat{p}}$ theory and will be used for the discussion of band dispersion near the zone center.

\subsection{\label{sec:NFE} Nearly free electron model}

In order to understand the origins of energy dispersion in the true band structure, we begin with a symmetry analysis of the nearly free electron (NFE) band structure, which we show along with corresponding symmetries of the zone center states in Fig. \ref{fig:lattice}(d). 

The lowest-energy state at the $\Gamma$-point (green hexagram) is of $\Gamma_1^+$ symmetry since its wavevector is the Brillouin zone origin, labeled [00]. The first excited states (blue square) are sixfold degenerate, with their wavevectors corresponding to the six reciprocal lattice points nearest to the origin [Fig. \ref{fig:lattice}(c)] and denoted by [01]. By examining their characters under all the $D_{3h}$ symmetry operators, it is easily found that these first excited states are composed of $2\Gamma_3^+\oplus 2\Gamma_1^+$. In fact, all reciprocal lattice points sitting on one of the three axes of in-plane $180^\circ$ rotation operators possess this symmetry. All other reciprocal lattice vectors share the same symmetry with the {\em second} excited states, denoted by red circles in Fig. \ref{fig:lattice}(d) and labeled [11] (the wavevector sum of any two neighboring [01] states). These are also six-fold degenerate and have the symmetry of $2\Gamma_3^+\oplus\Gamma_2^+\oplus \Gamma_1^+$. Given the fact that we are exclusively considering the low-energy bands, only the [00] and [01] states are of interest (the energy of [11] planewaves is nearly 30 eV). 

In the 2D NFE model, planewave eigenstates are always even under the in-plane mirror reflection operator $\sigma_h$. In a physically real lattice where the configurations of atomic orbitals have even or odd parities with respect to $\sigma_h$, the $\Gamma_3^+$ states of [01] can transform into $\Gamma_3^-$, while the $\Gamma_1^+$ states of both [01] and [00] can transform into $\Gamma_2^-$ (see the sign of characters for $\mathbb{C}_{3-5}$ that take $z\rightarrow-z$ in Table~\ref{tab:CharacterTable}). 

The true crystal lattice potential modifies this planewave dispersion while maintaining the symmetry. First of all, the finite thickness of the real monolayer (analogous to confinement in a quantum well) results in a series of subbands that originate from the primordial 2D states. Secondly, the in-plane potential breaks the $\Gamma$-point degeneracy of different IRs in the NFE band structure, and mixes the eigenstates into a linear combination of planewaves belonging to the same IR.  Numerical procedures utilizing planewaves as basis functions, such as the empirical pseudopotential method\cite{Chelikowsky_PRB76} and density-functional theory (DFT) packages like {\sc Quantum ESPRESSO},\cite{QE-2009} can be used to examine these planewave origins.

Our last remark about this NFE model is that the momentum matrix elements between reciprocal lattice planewaves -- even if they are allowed by symmetry -- are nonvanishing only for degenerate symmetrized planewaves. For example, Table~\ref{tab:CharacterTable} indicates that the momentum operator polar vector components $\hat{\pi}_x$ and $\hat{\pi}_y$ (we use this notation to avoid confusion with the $p_{x,y,z}$ orbitals) belong to the IR $\Gamma_3^+$, and their coupling between states belonging to $\Gamma_1^+$ and $\Gamma_3^+$ is apparently allowed by symmetry.  Momentum matrix elements between states of these representations, if both are composed of [01] symmetrized planewaves, clearly have magnitude on the order of $2\pi\hbar/a$ (where $a$ is the lattice constant). However, the coupling between $\Gamma_1^+$ of [00] and $\Gamma_3^+$ of [01] by $\hat{\pi}_{x,y}$ vanishes due to the structure of their oscillatory wavefunctions. We will show in the following subsection that this fact is essential in determining the band dispersion, because if two bands are coupled by the $\mathbf{k}\cdot\mathbf{\hat{p}}$ perturbation, they tend to energetically repel each other. The coupling strength, and hence quantities like effective masses, thus sensitively depends on whether the vestigial origin of these bands is the same planewave state in the NFE model. 

\subsection{\label{sec:mexhat} Origin of valence band distortion, \\{\textit{alias}} the `caldera'}

\begin{figure}
\includegraphics[scale=0.4]{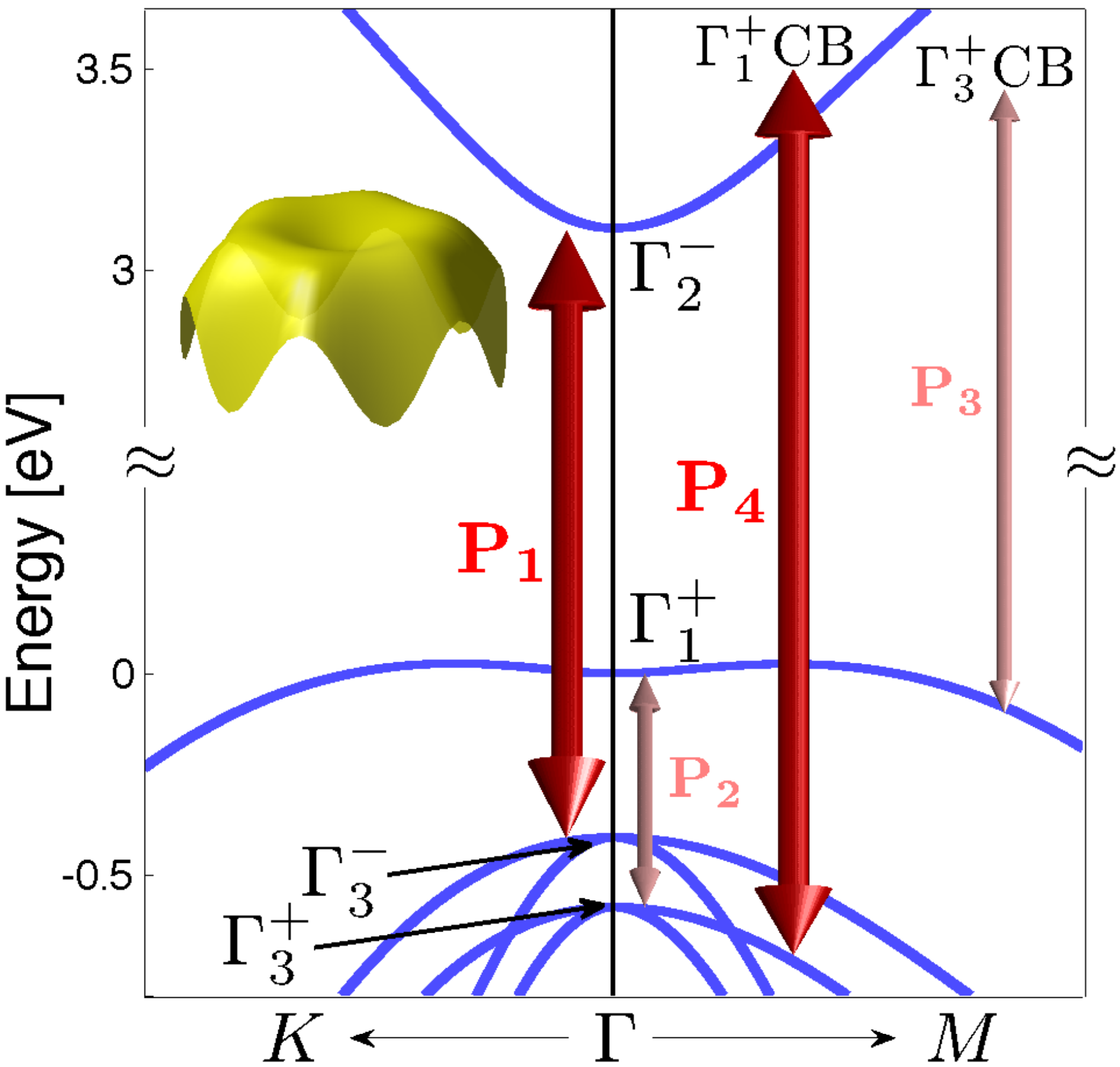}
\caption{Spin-independent tight-binding bandstructure around the zone center along the $K-\Gamma-M$ directions, in the vicinity of the bandgap. Here the $y$-axis (energy) is contracted within the gap region for better illustration. IRs of zone center eigenstates are shown. The double-headed arrows indicate dominant $\mathbf{k}\cdot\mathbf{\hat{p}}$ interactions responsible for a `caldera'-shaped highest valence band depicted by the 3D illustration inset.\label{fig:band_no_spin}}
\end{figure}

The empirical tight binding band structure of GaSe in the vicinity of the band gap is shown in Fig. \ref{fig:band_no_spin}. Besides the lowest $\Gamma_2^-$ conduction band, five of the nine valence bands are plotted, including the non-degenerate highest valence band $\Gamma_1^+$ and the two pairs of doubly-degenerate $\Gamma_3^\pm$. It is interesting to note that the remaining four lowest valence bands not shown in Fig. \ref{fig:band_no_spin} are composed of two pairs of $\{\Gamma_1^+,\Gamma_2^-\}$ (see, for example, Fig. 1 of Ref.~[\onlinecite{Robertson_JPC79}]). These states, differing by their definite parity with respect to $\sigma_h$, are each split by the sublayer bonding/antibonding energy. This heuristic can be confirmed by correlating the band splitting with the amplitude of $p_z$ orbital components (responsible for the strong inter-sublayer $\sigma$ bonding) in each pair. Likewise, the splitting of $\Gamma_3^\pm$ states is determined by the $\pi$ bonding/antibonding energy difference between the even and odd configurations of purely $p_x$ and $p_y$ wavefunctions.  

Our first-principles calculation using norm-conserving pseudopotentials shows that the $\Gamma_1^+$ highest valence band, together with the remaining four nondegenerate lowest valence bands just mentioned, originate from the NFE [00] planewave. Their common origin can be understood in terms of subband formation induced by electron confinement to the quasi-2D atomic lattice with finite thickness, as discussed above in Sec. \ref{sec:NFE}. For example, in the $\Gamma_1^+$ highest valence state, this [00] component accounts for $\approx 2/3$ of the wavefunction amplitude, while only 30\% comes from the [01] planewave. On the other hand, the two pairs of $\Gamma_3^\pm$ valence bands, as well as the lowest conduction band $\Gamma_2^-$, originate from the [01] planewave (more than 90\% in $\Gamma_3^\pm$ and 80\% in $\Gamma_2^-$).

In light of these wavefunction compositions, we can conclude from the argument at the end of the previous subsection that symmetry-allowed momentum matrix elements coupling these bands, represented in Fig. \ref{fig:band_no_spin} by
\begin{align}
&P_1 = \frac{\hbar}{m_0}\langle\Gamma_2^-|\hat{\pi}_{x,y} |\Gamma_3^-\rangle,\qquad\text{(red arrow)}\label{eq:P1}\\
&P_2 = \frac{\hbar}{m_0}\langle\Gamma_1^+|\hat{\pi}_{x,y} |\Gamma_3^+\rangle,\qquad\text{(short pink arrow)}
\label{eq:P2}
\end{align}
are quite different in magnitude. Specifically, the probability associated with the former ($|P_1|^2$) is about an order of magnitude larger than that of the latter ($|P_2|^2$, $\approx 0.3^2$ smaller). Therefore, the effective mass $m^*$ of $\Gamma_2^-$ is much smaller than the free electron mass $m_0$ (similar to the situation of the $\Gamma$-point conduction band minimum in many cubic systems such as GaAs and Ge), while $m^*$ of the $\Gamma_1^+$ valence band is slightly smaller than $m_0$ due to the weak upward repulsion from $\Gamma_3^+$ -- but still positive! 

This unusual dispersion (which would otherwise result in a smaller, or closed, bandgap) is counteracted by the interplay of several secondary factors. First of all, downward repulsion comes from upper conduction bands with $\Gamma_3^+$ symmetry indicated by $P_3$  in Fig. \ref{fig:band_no_spin} [with similar definition as Eq. (\ref{eq:P2})]. 
Our DFT calculation shows that the third-lowest conduction bands, with $\Gamma_3^+$ symmetry, originate from the [01] planewave. 
Close to the $\Gamma$-point, their downward repulsive influence on the $\Gamma_1^+$ highest valence band (again, dominated by the [00] planewave) cannot compete with the upward contribution from the close-by $\Gamma_3^+$ valence band. 
Further away from the $\Gamma$-point, however, this upward repulsion quickly vanishes, because the $\Gamma_3^+$ lower valence bands are strongly repelled downward by the dominant [01] planewave component of the $\Gamma_1^+$ second-lowest conduction band (indicated by $P_4$  in Fig.~\ref{fig:band_no_spin}). 
At large enough $k$, the $\Gamma_3^+$ conduction band wins over the valence band with the same symmetry, and overwhelms the (positive) quadratic free electron dispersion of the upper valence band. A large energy gap is thus opened.

In summary, the `caldera' shape of the upper valence band in the vicinity of the $\Gamma$-point is a result of the competition between the lower valence band and upper conduction bands with the same $\Gamma_3^+$ symmetry, similar to the origin of the possible indirect bandgap in phosphorene along the zigzag direction.\cite{Li_PRB14} The energy dispersion of the $\Gamma_1^+$ upper valence band can be analytically expressed via this $\mathbf{k\cdot\hat{p}}$ analysis and second-order perturbation theory by 
\begin{align}
E = \frac{\hbar^2k^2}{2m_0}
+\frac{|P_2|^2k^2}{E_A}
-\frac{|P_3|^2k^2}{E_B},
\label{eq:mex}
\end{align} 
where, in the second term,
\begin{align}
E_A = E_{1+}^{VB}-\left(E_{3+}^{VB}-\frac{|P_4|^2k^2}{E_{1+}^{CB}-E_{3+}^{VB}}\right)\label{eq:E_A}
\end{align} 
is the energy difference between the $\Gamma_1^+$ highest valence band and the $\Gamma_3^+$ valence band. The dispersion of the latter is taken into account by the last term of Eq.~(\ref{eq:E_A}), where $P_4$ has a definition similar to $P_2$ in Eq.~(\ref{eq:P2}) but with an amplitude that rivals $P_1$. 

Note that in the large-$k$ limit, the second term of Eq.~(\ref{eq:mex}) approaches a constant, and its contribution to energy dispersion vanishes. 
In the last term of Eq.~(\ref{eq:mex}), $P_3$ has similar amplitude to $P_2$, and the energy denominator
\begin{align}
E_B = \left(E_{3+}^{CB}-\beta k^2\right)-E_{1+}^{VB}\label{eq:E_B}
\end{align} 
is the difference between the $\Gamma_3^+$ conduction bands and the $\Gamma_1^+$ highest valence band. The $-\beta k^2<0$ term captures the negative effective mass of the $\Gamma_3^+$ conduction bands due to $\mathbf{k}\cdot\mathbf{\hat{p}}$ suppression from even further bands, and results in higher order dispersion of the $\Gamma_1^+$ valence band beyond $k^2$. Such higher order contributions, especially the quartic term, are important for valence band edge hole states  with relatively large wavevectors, e.g. at the caldera `rim'.

Given that $E_A$ is almost an order of magnitude smaller than $E_B$ at the $\Gamma$-point, it is very unlikely that the last term in Eq.~(\ref{eq:mex}) would overcome the sum of the first and the second terms at the zone center and render an ordinary parabolic hole band with negative $m^*$, as verified by both the empirical tight binding method and first-principles calculation. 
However, considering the possible deficiencies of numerical procedures (for example, underestimation of the bandgap in DFT), we cannot unequivocally assert the true nature of the valence band edge dispersion.
Furthermore, as we will show in the following section, such dispersion reversal of the highest valence band at the zone center is sensitive to the spin-orbit coupling strength.
Ultimately, the existence of this $\Gamma_1^+$ valence band `caldera' must be empirically verified by experiment, such as optical spectroscopy (see Sec. \ref{sec:opt_ori}) to detect the divergent DOS associated with high-order dispersion, or angle-resolved photoemission spectroscopy (ARPES) to directly probe the shape of the valence band.
 
With our fundamental understanding of the caldera's origin, we can also easily explain its gradual disappearance and transition into an ordinary parabolic valence band edge when monolayers are stacked and the system evolves toward bulk \cite{Li_SciRep14,Rybkovskiy_PRB14}. As the layer thickness increases, every band of the monolayer bandstructure develops a series of subbands corresponding to different van der Waals bonding with all possible interlayer phase configurations. The splitting of these subbands is therefore independent of all $\mathbf{k\cdot\hat{p}}$ interactions, and rather strongly relies on the wavefunction amplitude of vertically-distant chalcogen anion $p_z$ orbital component. The $\Gamma_1^+$ valence band edge has a large anion $p_z$ orbital component \cite{Robertson_JPC79} that diminishes away from the $\Gamma$-point, \cite{Rybkovskiy_PRB14} tending to raise the bottom of the caldera more than the rim. Furthermore, and just as important, this subband splitting changes the energy denominators in Eq. (\ref{eq:mex}), causing weaker (stronger) repulsion from lower (upper) $\Gamma_3^+$ bands. These altered interactions result in a gradually shallower caldera, which eventually disappears as bulk conditions are approached.

Lastly, we comment on the weak anisotropy of the valence band, depicted in the inset of Fig. \ref{fig:band_no_spin}. By taking into account even higher-order contributions beyond $k^4$, the circular extrema given by Eq.~(\ref{eq:mex}) at nonzero $k$ can be modulated by a term proportional to $k^6\cos6\phi$ (Ref.~[\onlinecite{Zolyomi_PRB13}]). 
The resulting six valence band maxima occur on the $\Gamma-K$ axes,  whereas saddle points exist (nearly equidistant from the zone center) on the $\Gamma-M$ axes. 

\section{\label{sec:symsoi} Symmetry \textit{avec} spin}

\subsection{Zeroth-order split-off states and spin mixing\label{sec:soizero}} 

\begin{figure}
\includegraphics[width=3in]{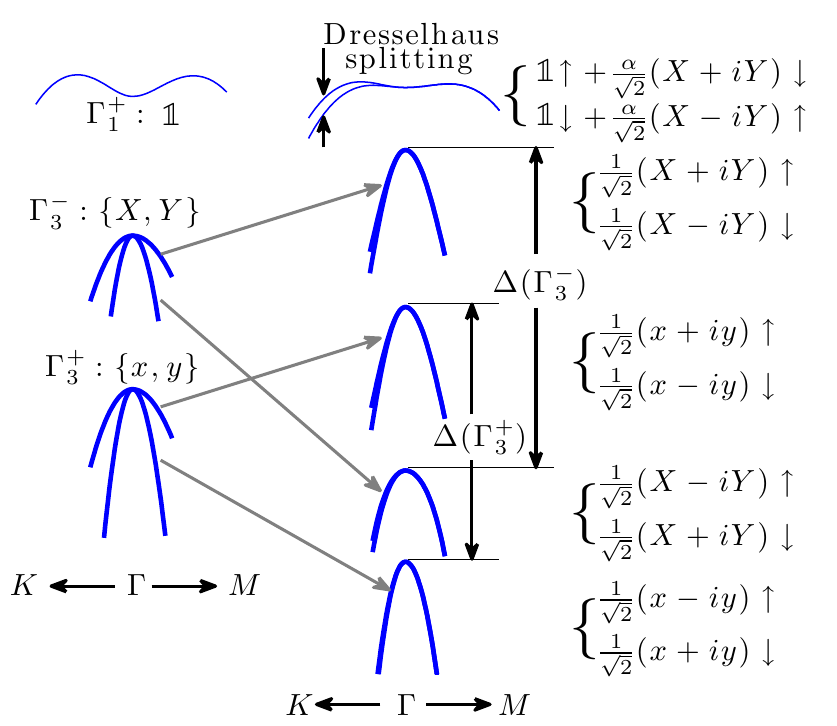}
\caption{Evolution of valence bands before (left) and after (right) introducing spin-orbit interaction. Gray arrows indicate spin-orbit induced broken degeneracy of the $\Gamma_3^\pm$ valence band. The symmetries of zone-center spin-dependent eigenstates are described by mixed basis functions given in Table~\ref{tab:CharacterTable}. Here, $X=yz$ and $Y=xz$.  \label{fig:band_spin}}
\end{figure}

The inclusion of spin-orbit coupling into the Hamiltonian can be treated within the framework of perturbation theory. Fig. \ref{fig:band_spin} depicts the important changes to the GaSe band structure among the five highest valence bands before (left) and after (right) spin-orbit interaction is considered. The spin-independent bands are identical to those in Fig. \ref{fig:band_no_spin}, except we have shortened the energetic distance between $\Gamma_1^+$ and $\Gamma_3^-$ for clearer illustration. Our incorporation of spin-orbit coupling into the tight-binding formalism follows the approach of Ref.~[\onlinecite{Chadi_PRB77}] by considering on-site spin-orbit parameters of Se and Ga atoms. 

The most significant hallmark of spin-orbit coupling is the splitting of the $\Gamma_3^\pm$ bands. Both of these bands have a double orbital degeneracy before including spin, similar to the top of the valence band in many cubic semiconductors in which the threefold degenerate valence band (sixfold if spin is included) is broken into the four-fold degenerate heavy- and light-hole bands and the doubly degenerate split-off hole bands. Here, taking $\Gamma_3^-$ as an example, the two orbital states can be denoted by the axial vector component basis functions as $\{X=yz,Y=xz\}$, see Table~\ref{tab:CharacterTable}. They are coupled by the zeroth-order spin-orbit perturbation $\frac{\hbar}{4m_0^2c^2}\nabla V\times\mathbf{\hat p}\cdot\vec{\sigma}$, so that the split-off energy is determined by the off-diagonal matrix element
\begin{align}
\Delta(\Gamma_3^-) = \frac{\hbar^2}
{2m_0^2c^2}\langle X|\left(\frac{\partial V}{\partial x} \frac{\partial}{\partial y}-\frac{\partial V}
{\partial y} \frac{\partial}{\partial x}\right)|Y\rangle,
\label{eq:Delta_p}
\end{align} 
and the spin-dependent eigenstates are a mixture of $X$ and $Y$ as shown in Fig. 3. The discussion for $\Gamma_3^+$ is the same, by replacing the basis functions $\{X,Y\}$ in Eq.~(\ref{eq:Delta_p}) with polar vector components $\{x,y\}$. Note that, due to the proximity of $\Gamma_3^+$ and $\Gamma_3^-$, this lowest-order spin-orbit interaction pushes down the lower pair of $\Gamma_3^-$ and raises up the upper pair of $\Gamma_3^+$ so much that their energies even switch order (crossing of the two gray arrows in the middle of Fig.~\ref{fig:band_spin}).

The lowest order spin-orbit interaction also couples the $\Gamma_3^-$ to the $\Gamma_1^+$ highest valence band, since the $\mathbf{k}$-independent spin-orbit invariants  $\{\sigma_x,\sigma_y\}$ belong to $\Gamma_3^- = \Gamma_3^-\otimes\Gamma_1^+$. This much weaker non-degenerate perturbation results in negligible energy shift compared with the split-off energy of $\Gamma_3^\pm$, but it does significantly alter the spin-dependent eigenstates by inducing spin-mixing. If we denote the spin-independent eigenstate of $\Gamma_1^+$ by the scalar basis function $\mathbb{1}$ as shown in Fig. \ref{fig:band_spin}, the unitless spin mixing coefficient $\alpha$ is
\begin{align}
\alpha= \frac{\hbar^2}{4m_0^2c^2 E_{1^+-3^-}} \langle\mathbb{1}|\left(\frac{\partial V}{\partial y} \frac{\partial}{\partial z}-\frac{\partial V}{\partial z} \frac{\partial}{\partial y}\right)|X\rangle.
\label{eq:alpha}
\end{align} 
Here, $E_{1^+-3^-}$ is the energy difference between $\Gamma_1^+$ and $\Gamma_3^-$, and their proximity (only several hundred meV) leads to a relatively large value of $\alpha$. Our tight-binding calculation shows that for $\Gamma_1^+$, the spin-mixing defined by the total square amplitude of the minority-spin components is $\alpha^2\approx 8\%$ at the $\Gamma$-point, falling only slightly to $\approx 5$\% at the caldera rim. For comparison, we also have an interaction between the $\Gamma_3^+$ valence bands and $\Gamma_2^-$ lowest conduction band that induces spin-mixing, yet it is much less pronounced due to the much larger energy denominator; our calculation gives $0.1\%$. Since the strength of Elliott-Yafet (EY) spin relaxation\cite{Elliott_PR54, Yafet_SSP63} is governed by the spin-mixing amplitude, the resulting EY relaxation rate for spin-polarized conduction electrons is nearly two orders of magnitude smaller than holes near the top of the valence band. Another important spin relaxation mechanism, Dyakonov-Perel (DP), is due to spin splitting of every band and is discussed in the next subsection (Sec. \ref{sec:Dressel}). 

Due to the mixing of $X$ and $Y$ components into the $\Gamma_1^+$ highest valence band by spin-orbit coupling, optical selection rules allow in-plane polarized electromagnetic radiation to connect states across the fundamental bandgap. This mixing is therefore especially important for experiments, since it allows carrier generation with normally-incident band-edge illumination.\cite{Gamarts_JETP77,Tang_PRB15}
We elaborate on radiative transitions and corresponding selection rules relevant for conduction-band spin polarization via optical orientation in Sec. \ref{sec:opt_ori}.

Lastly, we note that in Fig.~\ref{fig:band_spin} the zeroth-order spin-orbit coupling reduces the depth of the `caldera' distortion of $\Gamma_1^+$ highest valence band. As discussed in the previous section, this distortion is due to repulsive competition between the valence and conduction $\Gamma_3^+$ bands. Since the energy differences to $\Gamma_1^+$ are modified by the split-off energy $\Delta(\Gamma_3^+)$, dispersion close to the Brillouin zone center is sensitive to spin-orbit strength. Moreover, the spin mixing of $\Gamma_3^-$ components into the highest valence band allows its coupling to bands with $\Gamma_2^-$ symmetry via the $\mathbf{k}\cdot\mathbf{\hat{p}}$ perturbation,  incurring an additional downward repulsion contribution from the conduction band minimum. In fact, for the tight-binding GaSe model we use, positive effective mass at the $\Gamma$-point disappears for only $\approx$35\% stronger spin-orbit strength. With this concern, we reiterate our lack of absolute certainty on the question of  whether the quantitative nature of the highest valence band is a `caldera' shape or an ordinary hole-like paraboloid. 

\subsection{\texorpdfstring{$k^3$}{k-cubic} spin splitting: Dresselhaus effect\label{sec:Dressel}}

A remaining feature of the spin-dependent band structure shown in Fig.~\ref{fig:band_spin} is the splitting of opposite spin subbands for $\mathbf{k}$ along $\Gamma-K(K')$ directions. Although this splitting vanishes on the $\Gamma-M$ axes (the $C_{2v}$ group of the wavevector for points on these axes only has two-dimensional double group IRs), it generally persists for an arbitrary choice of $\mathbf{k}$. This splitting is similar to Dresselhaus spin-splitting in zincblende semiconductors,\cite{Dresselhaus_PR55} where it vanishes along $\Gamma-X$ \cite{Cardona_PRB88}. Likewise, it has the same fundamental cause, namely the absence of space inversion symmetry that allows a spin-orbit-induced effective internal magnetic field in reciprocal space. In this subsection, we analyze the symmetry of this higher-order spin-orbit effect using the method of invariants.\cite{BirPikus}

There are two potential approaches within this framework. One could extend single group to double group notation, in which the Pauli matrices of spin orbit interaction are embedded as invariant matrices, while $\mathbf{k}$-dependent terms are filled into the correct position in the matrices as invariant components.\cite{Song_PRB12} However, for simplicity we follow an alternative approach, considering both $\mathbf{k}$ and Pauli matrices (and their combinations) as invariant components within the single group notation.\cite{Li_PRL11}

We begin by taking the $\Gamma_1^+$ highest valence band as an example. The direct product of this IR with itself gives $\Gamma_1^+\otimes\Gamma_1^+ = \Gamma_1^+$, indicating that any term including $\sigma_x$ or $\sigma_y$ is forbidden. These two Pauli matrices behave as axial vectors (see Table~\ref{tab:CharacterTable}), which are odd under the operation of in-plane mirror reflection $\sigma_h$. The same is true for their combination with $k_x$ and $k_y$ (both are even under $\sigma_h$). This existence of the in-plane mirror reflection is a critical factor that determines the form of Dresselhaus spin splitting. In wurtzite systems sharing the threefold rotation symmetry but lacking $\sigma_h$ as a group element, $\sigma_x$ or $\sigma_y$ are allowed in the spin-dependent Hamiltonian and the lowest order Dresselhaus field is linear in $k_x$ and $k_y$.\cite{Rashba_FTT59, LYV_PRB96}

\begin{figure}
\includegraphics[scale=0.47]{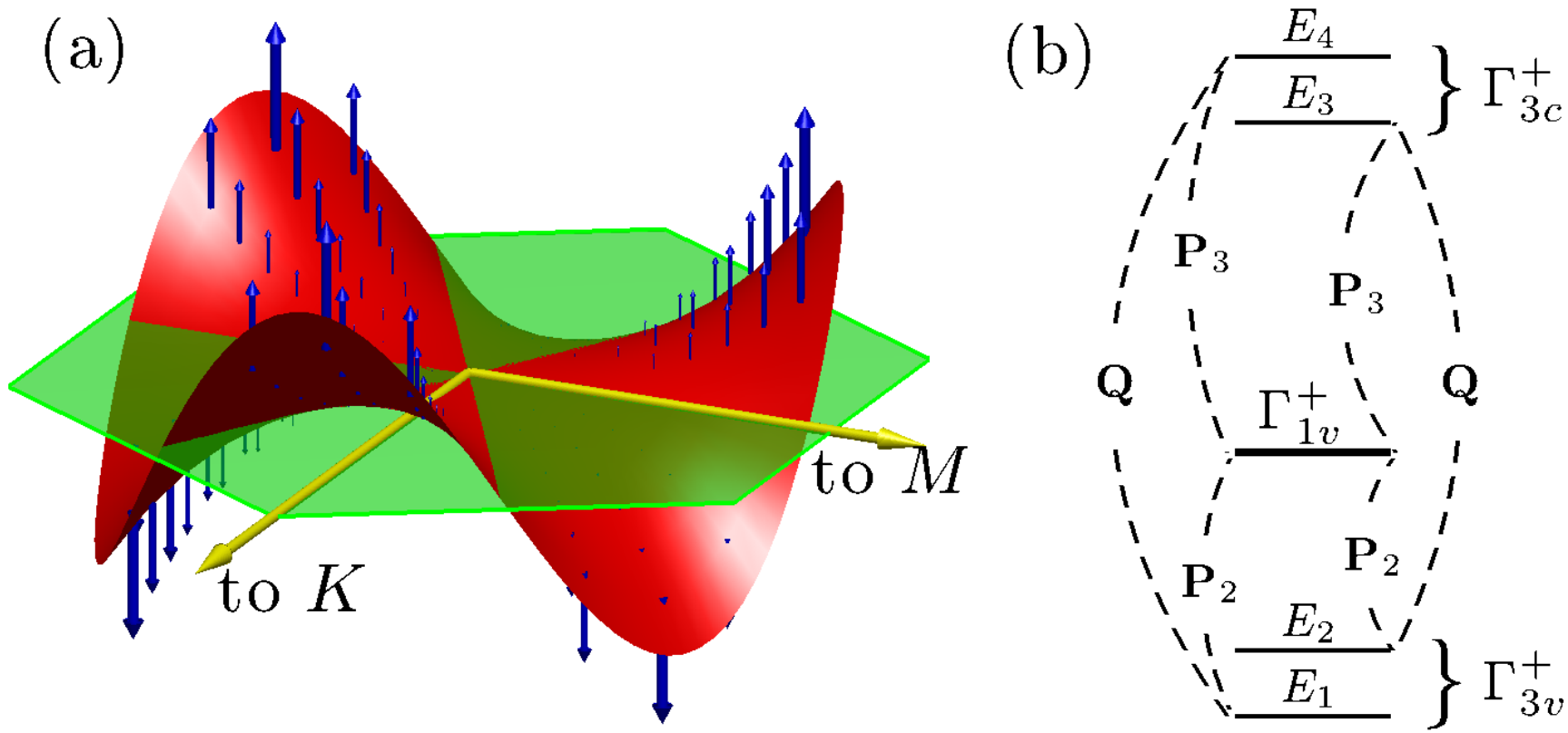}
\caption{(a) Spin-orbit-induced $k$-cubic Dresselhaus splitting [Eq.~(\ref{eq:Dresselhaus})] of the spin-up and spin-down subbands is represented by the red manifold. The green hexagonal plane is similar to the reduced Brillouin zone with $\Gamma-K$ and $\Gamma-M$ axes shown. Blue vectors represent the Dresselhaus internal magnetic field. Note that the out-of-plane field direction is reversed for $K$ and $K'$ and vanishes along $\Gamma-M$. (b) Dominant perturbation paths (the two dashed loops) for the calculation of the coefficient $\gamma_1$ in Eqs.~(\ref{eq:Dresselhaus}) and (\ref{eq:gamma1}). Horizontal lines represent the spin-dependent $\Gamma$-point states, with their single group origin listed on the right. $E_{1-4}$ are the energy values of the spin-split states with $\Gamma_3^+$ single group origin, with respect to the energy of $\Gamma_{1v}^+$, in ascending order. $P_2$ and $P_3$ are the same as they are in Fig.~\ref{fig:band_no_spin}, and $Q$ is the momentum matrix element between conduction and valence bands with the same $\Gamma_3^+$ symmetry.  \label{fig:dressel}}
\end{figure}

The remaining Pauli matrix, $\sigma_z$, belongs to $\Gamma_2^+$. Its combination with polynomials of $k_x=k\cos{\phi}$ and $k_y=k\sin{\phi}$ that also belong to $\Gamma_2^+$ are symmetry-allowed invariant components for the $\Gamma_1^+$ band ($\Gamma_2^+\otimes\Gamma_2^+=\Gamma_1^+$). We find that $\sin 3\phi=3\cos^2\phi\sin\phi-\sin^3\phi$ (requiring $k^3$ dependence) is the lowest-order invariant component that belongs to $\Gamma_2^+$ . With all these concerns, we conclude that the lowest-order Dresselhaus term can be written as
\begin{align}
H_{SO}^{\mathbf{k}}= \gamma_1 k^3 \sin 3\phi \,\sigma_z,
\label{eq:Dresselhaus}
\end{align}
where $\gamma_1$ is the spin-orbit coupling strength coefficient for this band. The form of Eq.~(\ref{eq:Dresselhaus}) is universal for all other bands (with different $\gamma$ coefficients), including those originating from the two-dimensional $\Gamma_3^{\pm}$ bands, due to the constraint imposed by $\sigma_h$ and three fold rotational symmetries. Fig.~\ref{fig:dressel}(a) is a schematic representation of Eq.~(\ref{eq:Dresselhaus}), where blue arrows represent the effective internal magnetic field. The field magnitude is proportional to $k^3$, and alternates direction between parallel and anti-parallel relative to the surface normal as a function of polar angle.

The spin-splitting parameter $\gamma_1$  in Eq.~(\ref{eq:Dresselhaus}) can be calculated using fourth-order perturbation theory. As in the corresponding calculation for III-V semiconductors, this quantity involves three matrix elements of the $\mathbf{k}\cdot\hat{\mathbf{p}}$ interaction and one of the spin-orbit coupling $\frac{\hbar}{4m_0^2c^2}\nabla V\times\mathbf{\hat p}\cdot\vec{\sigma}$. Alternatively, spin-orbit can be taken into account to all orders by using the exact split-off energies in third-order perturbation term denominators.\cite{Cardona_PRB88}

As discussed in the previous section, the $\mathbf{k}\cdot\hat{\mathbf{p}}$ terms belong to $\Gamma_3^+$ and can couple the $\Gamma_{1v}^+$ highest valence band only to bands with $\Gamma_{3}^+$ symmetry. The $\Gamma_{3}^+$ intermediate states, in turn, can also be coupled to each other by $\mathbf{k}\cdot\hat{\mathbf{p}}$, since $\Gamma_{3}^+\otimes\Gamma_{3}^+=\Gamma_{1}^+\oplus\Gamma_{2}^+\oplus\Gamma_{3}^+$. The dominant third-order  $\mathbf{k}\cdot\hat{\mathbf{p}}$ perturbation paths are therefore those shown in Fig.~\ref{fig:dressel}(b), and include the $\Gamma_{3v}^+$ valence band and the lowest $\Gamma_{3c}^+$ conduction band (the same as those given in Fig.~\ref{fig:band_no_spin}) as intermediate states.
For simplicity, their energies with respect to $\Gamma_{1v}^+$ in ascending order are denoted by $E_{1-4}$; clearly, $E_2-E_1$ is equal to $\Delta(\Gamma_3^+)$ given in Fig.~\ref{fig:band_spin}, and $E_4-E_3$ is the split-off energy of the conduction $\Gamma_{3c}^+$ states (with a similar definition). By summing over all paths in Fig.~\ref{fig:dressel}(b), one obtains
\begin{align}
\gamma_1&=\frac{\hbar^3}{m_0^3}\sum_{i,j = \Gamma_{3\!v,c}^+}\frac{\langle\mathbb{1}|\hat{\pi}_y|i\rangle\langle i|\hat{\pi}_y| j\rangle\langle j| \hat{\pi}_y|\mathbb{1}\rangle}{E_iE_j}\nonumber\\
&=|P_2QP_3|\left(\frac{1}{E_2E_3}-\frac{1}{E_1E_4}\right),
\label{eq:gamma1}
\end{align}
where $P_{2}$ and $P_{3}$ are given in Fig.~\ref{fig:band_no_spin} and Eq.~(\ref{eq:P2}), and $Q$ is the $\mathbf{k}\cdot\hat{\mathbf{p}}$ matrix element $\frac{\hbar}{m_0}\langle\Gamma_{3v}^+|\hat{\pi}_{x,y} |\Gamma_{3c}^+\rangle$. 
Due to the partitioning of $x$ and $y$ components in spin-orbit-split $\Gamma_3^+$ wavefunctions (as shown in  Fig.~\ref{fig:band_spin}), $Q$ couples energy levels only between $E_1$ and $E_4$, and between $E_2$ and $E_3$,
\footnote{More detailed explanation requires the double group framework, in which the single group $\Gamma_3^+$ IR evolves into two double group IRs as: $\Gamma_3^+\otimes D_{1/2} = \Gamma_8\oplus \Gamma_{9}$ in the Bethe notation (the spinor $D_{1/2}$ is denoted by $\Gamma_7$). $\Gamma_8$ corresponds to states of $E_1$ and $E_3$ in Fig.\ref{fig:dressel}(b), while $\Gamma_9$ corresponds to those with energies $E_2$ and $E_4$. States belonging to $\Gamma_8$ and $\Gamma_9$ can be coupled by 
$\mathbf{k}\cdot\hat{\mathbf{p}}$ since $\Gamma_8\otimes \Gamma_9=\Gamma_3^+\oplus\Gamma_3^-$ includes $\Gamma_3^+$. However, $\mathbf{k}\cdot\hat{\mathbf{p}}$ cannot couple $\Gamma_8$ or $\Gamma_9$ to themselves, since $\Gamma_3^+$ is neither included in $\Gamma_8\otimes \Gamma_8=\Gamma_1^+\oplus\Gamma_2^+\oplus\Gamma_1^-\oplus\Gamma_2^-$ nor $\Gamma_9\otimes \Gamma_9=\Gamma_1^+\oplus\Gamma_2^+\oplus\Gamma_3^-$.} 
so that in Eq.~(\ref{eq:gamma1}) there are no terms with  energy denominator $E_1 E_3$ nor $E_2 E_4$.
Note that the corresponding calculation for the $\Gamma_2^-$ lowest conduction band involves dominant contribution from the $\Gamma_{3v,c}^-$ bands, and is similarly dependent on the spin-orbit-induced splitting $\Delta(\Gamma_3^-)$.

The out-of-plane orientation of the internal magnetic field in group-III metal-monochalcogenides resembles the case of III-V zincblende semiconductor [110] quantum wells, except that in the latter system Dresselhaus splitting is linear in $\mathbf{k}$ (see Ref.~[\onlinecite{Winkler_PRB04}]). In that case, spins oriented normal to the plane have relatively much longer lifetime than in-plane spins because they are eigenstates in the Dresselhaus field, and thus the Dyakonov-Perel (DP) spin relaxation mechanism vanishes.\cite{Dyakonov_SPSS1972} We therefore expect strong spin relaxation anisotropy in the monochalcogenides for the same reason. Specifically, when the spin orientation is chosen to be in-plane, spin relaxation is governed by DP and proportional to $\gamma_1$. On the other hand, spin with out-of-plane orientation is not subject to precession in the Dresselhaus field and DP is absent; in that case, EY (determined by the spin-mixing coefficient $\alpha$) then dominates spin relaxation.

Another feature of the [110] zincblende quantum well, enabled by the $\mathbf{k}$-linear spin splitting and $k$-quadratic band dispersion, is the possibility to generate a so-called `persistent spin helix'.\cite{Schliemann_PRL03, Koralek_Nature09, Bernevig_PRL06} This static spin texture results from the cancellation of $k$-dependence in the ratio of spin-orbit field to group velocity determining precession angle. The monochalcogenide hole states at the $\Gamma_1^+$ valence band maximum have a large $k^4$ dispersion component, and together with $k^3$ Dresselhaus splitting thus can also support a `spin helix'. However, the underlying threefold rotation symmetry eliminates its `persistence' in the presence of scattering.

As long as the threefold rotational symmetry is captured, terms higher than cubic in $\mathbf{k}$ are allowed, if constructed by combining Eq.~(\ref{eq:Dresselhaus}) with an additional invariant component belonging to $\Gamma_1^+$. For example, noting that $k^2$ belongs to $\Gamma_1^+$, the possible presence of fifth-order, seventh-order, etc. in $\mathbf{k}$ cannot be ruled out. These higher-order contributions become prominent at large $k$, especially for bandedge states on the rim of the valence band `caldera'. If such higher order terms have the opposite sign as the lowest-order cubic term, the Dresselhaus spin splitting would appear sub-cubic as $k$ increases. However, we must emphasize that in this system, {\em a linear Dresselhaus term is forbidden by symmetry}. Any contrived attempt to fit a numerically-calculated spin splitting using only linear and cubic terms, as from DFT in Ref.~[\onlinecite{Do_arxiv15}], is physically unjustified and susceptible to misleading evidence. Mistaken conclusions are an inevitable consequence of artificial symmetry breaking stemming from e.g. numerical rounding errors (associated with unavoidably representing the irrational values of atomic position components of the hexagonal Bravais lattice in a finite-digit scheme) at some arbitrary level of precision.

\section{\label{sec:gfactor} Orbital magnetism and effective \lowercase{g}-factor}
As previously mentioned, the spin-orbit-induced splitting of lower $\Gamma_3^\pm$ valence bands is reminiscent of the $\Gamma_4\rightarrow \Gamma_7 \oplus \Gamma_8$ valence band splitting in bulk cubic semiconductors. In that familiar case, the valence-band splitting allows a nonzero orbital contribution to the magnetic moment and a conduction-band g-factor substantially different from the spin-only value of 2, with smaller band gap and larger split-off energy strongly enhancing the correction.\cite{Roth_PR59, Graf_PRB95, Yafet_SSP63} Heuristically, we can surmise that large orbital g-factor in a given band follows from coupling to nearby bands which originate from states whose initial orbital degeneracy has been split by spin-orbit interaction.  In $MX$, the lower $\Gamma_3^\pm$ valence bands are indeed split by an amount on the order of the band separation to the upper $\Gamma_1^+$ valence band, suggesting a large orbital g-factor there. To test this assumption, we therefore undertake the appropriate calculation below.

In a magnetic field, we make the Peierls substitution by writing the canonical momentum vector $\mathbf{\hat{\pi}}\rightarrow \mathbf{\hat{\pi}}+e\mathbf{A}$. An out-of-plane field $B_z\mathbf{z}$ has vector potential $\mathbf{A}=B_zx\mathbf{y}$ (in the Landau gauge) so that the Hamiltonian term giving rise to orbital diamagnetism is $\frac{e\hbar}{4m}(4\frac{\hat{\pi}_y\hat{x}}{\hbar})B_z$. The expectation value of the operator in parenthesis can be evaluated in the Bloch band basis to yield the orbital magnetic moment (in units of $\frac{e\hbar}{4m}=\frac{\mu_B}{2}$) that competes with the spin magnetic moment to give an overall effective g-factor $<+2$.

\begin{figure}
\includegraphics[width=3.5in]{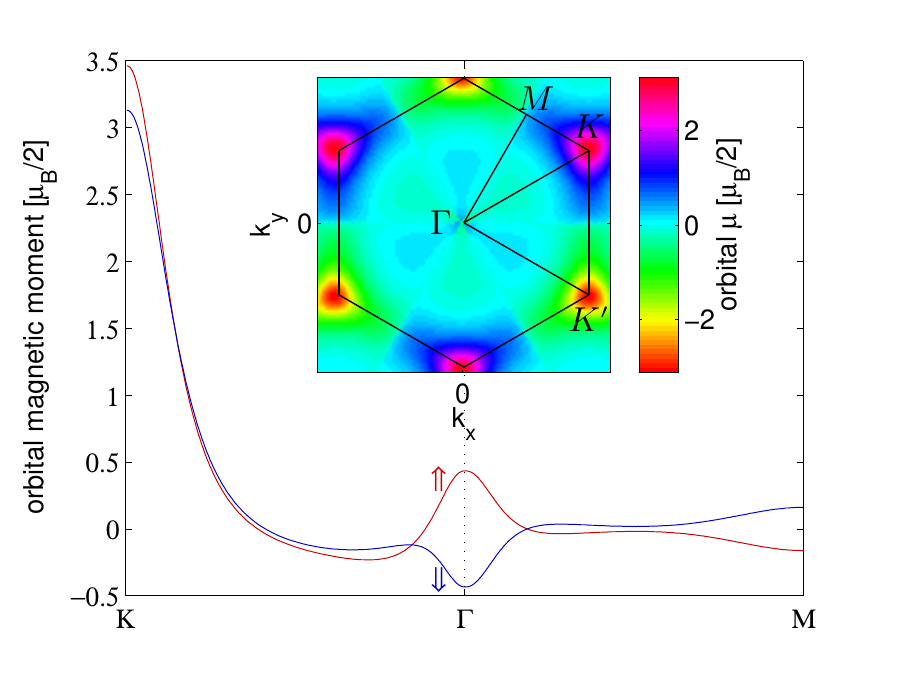}
\caption{$K-\Gamma-M$ orbital magnetic moment of both spin states in the $\Gamma_1^+$ valence band. Units chosen indicate orbital contribution to effective g-factor. Inset: Orbital magnetic moment of uppermost valence band within the full Brillouin zone, highlighting $K/K'$ valley magnetism. \label{fig:gfactor} }
\end{figure}

Unfortunately, the position operator $\hat{x}$ cannot be evaluated directly due to the indeterminate nature of matrix-element integrals over delocalized and non-normalizable Bloch waves that extend to infinity. \cite{Resta_PRL98} Instead, using the Ehrenfest theorem $\langle \hat{\pi}_x\rangle =mv_x=m\frac{d\langle x\rangle}{dt}=\frac{1}{i\hbar}[x,\mathcal{H}]$ and taking matrix elements in a band basis $|\psi_n\rangle$, we obtain $\langle \psi_n | \hat{x} |\psi_{n'} \rangle =\frac{i\hbar}{m}\frac{\langle \psi_n | \hat{\pi}_x |\psi_{n'} \rangle}{E_n-E_{n'}}$, as in Ref.~[\onlinecite{Adams_JCP53}]. Thus, the orbital ``g-factors" of each band can be evaluated using the operator with matrix elements

\begin{equation}
\langle n | g_{orbit}|n''\rangle=\frac{4i}{m}\sum_{n'\neq n}\frac{\langle \psi_n | \hat{\pi}_x |\psi_{n'} \rangle\langle \psi_{n'} | \hat{\pi}_y |\psi_{n''} \rangle}{E_n-E_{n'}}.\label{eq:gfactor}
\end{equation}

\noindent In a symmetric gauge, $\hat{\pi}_y\hat{x}\rightarrow \frac{1}{2}(\hat{\pi}_y\hat{x}-\hat{\pi}_x\hat{y})$, with appropriate change to the numerator in Eq. \ref{eq:gfactor}. We note that in the tight-binding formalism it is straightforward to use momentum operators constructed directly from the Hamiltonian $\mathcal{H}$, via $\mathbf{\hat{\pi}}=\frac{m}{\hbar}\vec{\nabla}_k \mathcal{H}$, although this ignores intra-atomic orbital overlap contributions while maintaining gauge invariance\cite{Pedersen_PRB01, Foreman_PRB02, Boykin_PRB01}.

Immediately we see that the coupling to bands governing the orbital g-factor is through matrix elements of orthogonal components of momentum. [An interesting example is phosphorene,\cite{Li_PRB14} where due to strongly broken symmetry all orbital bands are nondegenerate, and furthermore along the high-symmetry axes in $k$-space the momentum matrix elements of $\hat{\pi}_x$ and $\hat{\pi}_y$ are never simultaneously nonzero, so that orbital g-factor of every band vanishes.] According to Table~\ref{tab:CharacterTable}, the polar vector components $\hat{\pi}_{x,y}$ transform like $\Gamma_3^+$, so coupling to the $\Gamma_1^+$ upper valence band is dominated by the $\Gamma_3^+$ lower valence band (and $\Gamma_3^+$ upper conduction bands, but to a lesser degree due to the large energy denominator).

As discussed in Sec. \ref{sec:Dressel}, the spin-orbit field vanishes along $\Gamma-M$ and spin degeneracy is preserved. For this axis, orbital magnetic moment for spin up and spin down in the $\Gamma_1^+$ upper valence band, calculated using first-order degenerate perturbation theory to evaluate Eq.~(\ref{eq:gfactor}), are equal in magnitude but opposite in sign as expected (see Fig.~\ref{fig:gfactor}). However, lattice inversion asymmetry allows valley-dependent magnetic interactions,\cite{Yao_PRB08, Xu_NatPhys14} and this is manifest in the behavior of orbital magnetic moment along the $\Gamma-K$ axis, where spin degeneracy is lifted. Instead of sharing equal magnitude orbital magnetic moment with a nearly degenerate state at the same $k$, here the opposite magnetic moment is found with its time-reversed partner (along $\Gamma-K'$). Thus, all three $K$ points (oriented at $120^\circ$ in $k$-space) have identical magnetic moment (and at $|g|\approx 3$, several times larger than at $\Gamma$), with all three complementary $K'$ points having the opposite moment, nearly independent of spin (see inset to Fig. \ref{fig:gfactor} for g-factor in the full BZ). The lowest conduction band shows a similar dichotomy between behavior of magnetic moment along $\Gamma-M$ and $\Gamma-K$, but the orbital g-factor for electrons is quite small, reaching only approximately $|\langle g_{orbit}\rangle|=$0.3 (not shown).

This relatively small orbital correction to effective g-factor is contrary to our initial heuristic expectation of a large correction in the $\Gamma^+_1$ band based on spin-orbit splitting of the nearby $\Gamma_3^+$ band. However, while the denominators in Eq. \ref{eq:gfactor} are indeed small and unequal as expected, the momentum matrix elements between these two bands is weak. This can be understood by recalling from Sec. \ref{sec:NFE} that the dominant planewave origin of the $\Gamma_1^+$ upper valence band is from [00], whereas the lower $\Gamma_3^\pm$ valence bands are rooted instead in the [01] planewaves [see Fig. \ref{fig:lattice}(d)]. Momentum matrix elements between these two bands, dominated by contributions from different planewave origins, are inherently weak.

\section{\label{sec:opt_ori} Optical orientation}

With appropriate selection rules and spin-orbit splitting of the otherwise degenerate valence band, conduction electron spin polarization can be generated via interband dipole excitation.\cite{Lampel_PRL68} This ``optical orientation" process is most clear in cubic III-V semiconductors with a ($\ell=1$) $p$-like valence band, where one $j=1/2$ band is split off from the remaining $j=3/2$ states by lowering an energy $\Delta$. Then, circularly-polarized photons with energy $\hbar\omega$ at the bandgap $E_g$ can optically excite electrons into the conduction band, with matrix element asymmetry for spin up to spin down at the zone center of 3:1. This ratio is determined by the orbital wavefunction components (e.g. Clebsch-Gordan coefficients), giving spin polarization $P=\frac{3-1}{3+1}=50\%$ for $\hbar\omega=E_g$.  \cite{Dyakonov_SPJETP71, *Dyakonov_ZhETF71} Higher spin polarization can be obtained in 2D epitaxial quantum wells made from the same materials, where confinement induces splitting of the light and and heavy hole states.\cite{Uenoyama_PRL90, Pfalz_PRB05}

Optical orientation in bulk GaSe was explored both theoretically\cite{Ivchenko_JETP77, *Ivchenko_ZhETF77} and experimentally\cite{Gamarts_JETP77, *Gamarts_ZhETF77} in the 1970s (and more recently in Ref.~[\onlinecite{Tang_PRB15}]), but the symmetry of single-layer GaSe is expected to yield different results.

\subsection{\label{sec:absorb} Absorption Spectroscopy}

We calculate the corresponding optical orientation in the conduction band of monolayer GaSe within the tight-binding framework, by sampling wavevectors within the full Brillouin zone and compiling the squared (circularly-polarized) optical matrix elements in 
\begin{align}
\!\!\!A^{\uparrow/\downarrow}(\hbar\omega)\propto
\sum_{n\neq c}|\langle\psi_c^{\uparrow/\downarrow}|\hat{\pi}_\pm |\psi_n\rangle|^2\delta(\hbar\omega-(E_c-E_n)),
\end{align}
where $\hat{\pi}_\pm=\hat{\pi}_x\pm i\hat{\pi}_y$ determines the chirality of the circularly-polarized electromagnetic field; this approach restricts our analysis to direct transitions only. We subsequently calculate the conduction spin polarization via $P_c=\frac{A^\uparrow-A^\downarrow}{A^\uparrow+A^\downarrow}$.

\begin{figure}
\includegraphics[width=3.25in]{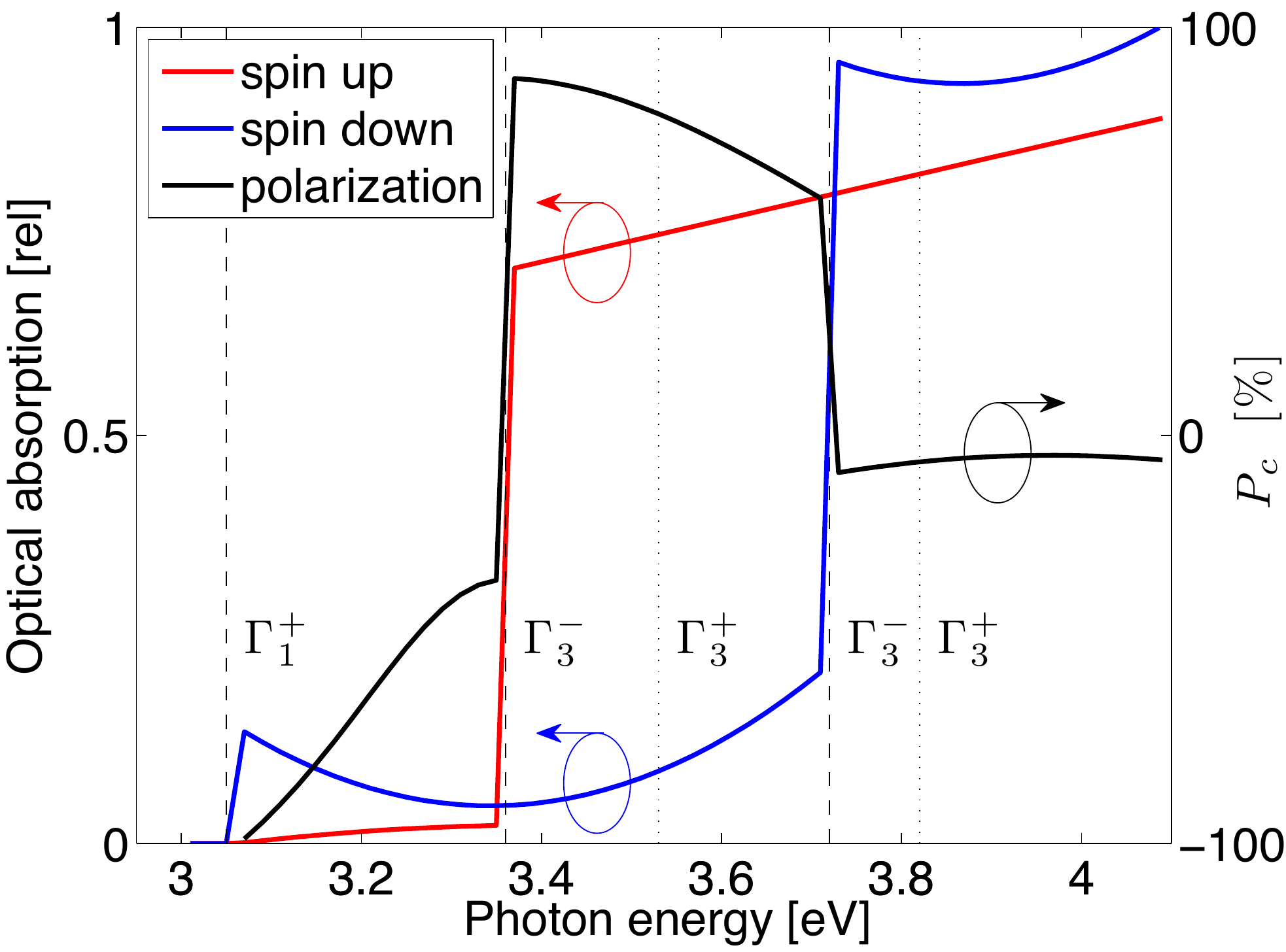}
\caption{Spin-resolved optical absorption of right-handed circular polarized light resulting in excitation of electrons into the $\Gamma_2^-$ conduction band, and resulting spin polarization (black line, right axis). Critical band edge energies are indicated by dashed or dotted lines. \label{fig:optori} }
\end{figure}

The results from a sample of $10^6$ pseudo-random $k$ points is shown in Fig.~\ref{fig:optori}, where transition probability via right-handed circularly polarized light into the spin up (down) conduction band is shown in red (blue), together with the spin polarization $P_c$ in black. One striking feature of these curves is the abrupt step-like absorption edges due to the 2D DOS singularity caused by the `caldera' valence dispersion. Because of this behavior, we are justified in displaying the least-squares fit to the spectroscopy with a piecewise quadratic function. Note that the divergent $\propto(-E)^{-1/2}$ density of states associated with the quartic $\propto k^4$ dispersion in the $\Gamma_1^+$ valence band causes the near-band-edge absorption to initially fall as photon energy $\hbar\omega$ increases. The existence of the putative `caldera' valence dispersion can therefore be straightforwardly supported by this one experimental signature, if it can be isolated from excitonic modification to the absorption spectrum.

Due to dominant $\Gamma_1^+$ character of the upper valence band, optical absorption of normally-incident circularly polarized photons at the fundamental band edge is suppressed (i.e. $\langle\Gamma_1^+|\Gamma_3^+|\Gamma_2^-\rangle=0$). However, the small quantity of $\frac{\alpha}{\sqrt{2}}(X\pm i Y)$ spatial character with opposite spin mixed into the wavefunction from the $\Gamma_3^-$ lower valence band (see Fig.~\ref{fig:band_spin}) results in perfect -100\% optical orientation, despite weak optical efficiency.   

Once the photon energy $\hbar\omega$ is sufficient to excite electrons from the $\Gamma_3^-$ band at approximately 3.35~eV, the conduction band spin polarization nearly fully reverses due to the much stronger optical matrix element from a dominant $\frac{1}{\sqrt{2}}(X\pm iY)$ character. However, further increases in energy by the split-off increment $\Delta(\Gamma_3^-)$ populate both spin states nearly equally, destroying the optical orientation for $\hbar\omega\apprge$3.7~eV. Again, this transition is much more abrupt than in the case of cubic zincblende direct-gap semiconductors because of the higher zone-center density of states in each band.

\subsection{Valence band polarization via relaxation\label{sec:spinrelax}}

\begin{figure}
\includegraphics[width=3in, height=8in]{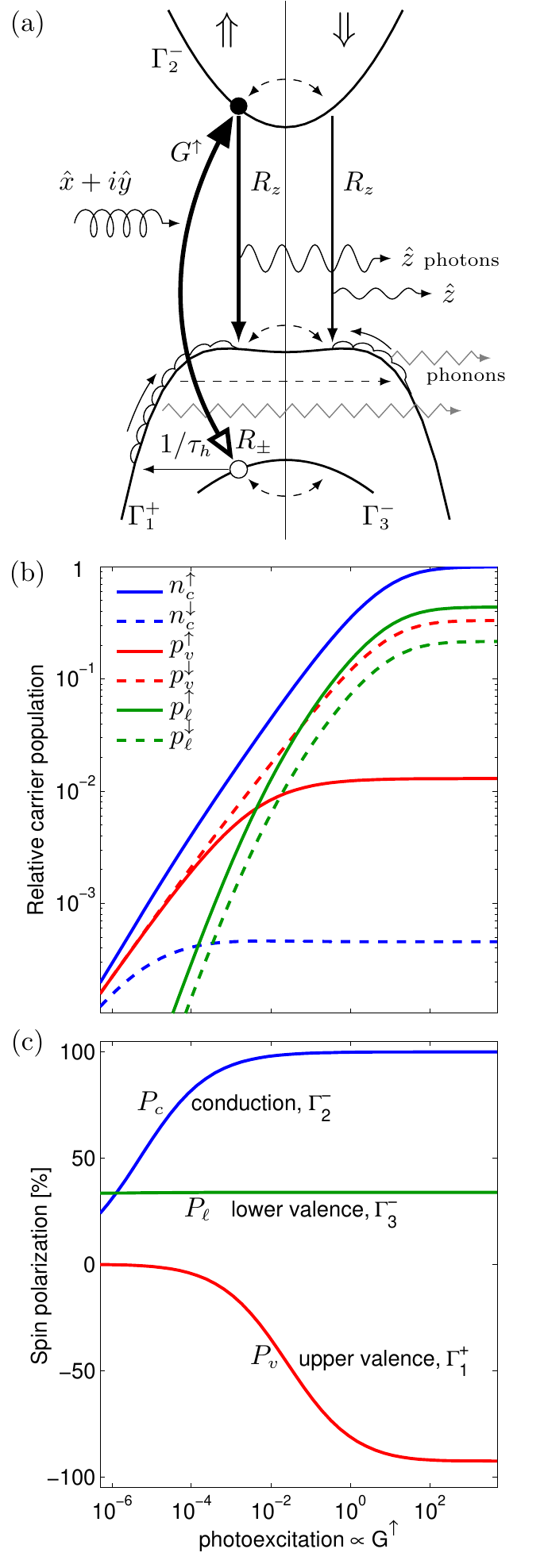}
\caption{(a) Schematic illustration showing optical orientation of spin-polarized conduction electrons via circularly-polarized electromagnetic excitation from the lower valence band and subsequent relaxation dynamics resulting in polarization of the upper valence band. (b) Spin-dependent carrier density in an equivalent three-level system modeled by Eqs. \ref{eq:rateeqns_first}-\ref{eq:rateeqns_last}, and (c) the corresponding spin polarizations. In normalized units, parameters are $\tau_c=80$, $\tau_v=1$, $\tau_\ell=1$, $R_z=10$, $\tau_h=1$, $D_c=5$, $D_\ell=5$, and $G^\downarrow=0$. 
\label{fig:holepol}}
\end{figure}

An intriguing phenomenon is enabled by the spin structure of electron states and corresponding selection rules in these monochalcogenides. Refer to Fig. \ref{fig:holepol}(a):
Optical excitation with normally-incident circularly polarized light of sufficient energy generates spin polarized carriers in the $\Gamma_2^-$ conduction and the upper branch of the spin-split $\Gamma_3^-$ lower valence band. 
Direct transition of the $\Gamma_3^-$ hole to the $\Gamma_1^+$ upper valence band is suppressed by the optical selection rules involving dominant character of the wavefunctions of each band, so low-energy but large-momentum phonon emission is necessary for relaxation. This scattering event takes an initially spin-polarized hole near the $\Gamma$-point to states far away from the zone center, from where it can relax to the valence band maximum via a cascade of optical phonon emission events, shedding tens of meV energy at each step.\cite{Bashenov_PSS78, Altshul_PSS80}
Due to strong spin-orbit-induced spin mixing (with amplitude $\alpha$, see Fig. \ref{fig:band_spin}) in this band, the Elliott spin scattering mechanism results in virtually complete depolarization upon thermalization. However, the $\Gamma_2^-$ conduction band and $\Gamma_1^+$ valence band are connected by a strong $\hat{\pi}_z$ optical matrix element with $\Gamma_2^-$ symmetry, so radiative relaxation proceeds via linearly-polarized luminescence emitted in an in-plane direction. Since this direct interband transition is spin-conserving, oppositely polarized holes will remain in the $\Gamma_1^+$ valence band.  

This dynamic upper valence band hole polarization can be modeled via a rate-equation approach, similar to the simulation of recent time-resolved luminescence studies in thin but still 3D GaSe flakes. \cite{Tang_JAP15,Tang_PRB15} A simplified equivalent three-level system is shown in Fig. \ref{fig:holepol}(a), where spin-conserving interband transitions ($\Gamma_3^-\leftrightarrow\Gamma_2^-$ optical orientation with generation rate constant $G^{\uparrow}$ and spontaneous relaxation rate constant $R_\pm$, and $\Gamma_1^+\rightarrow\Gamma_2^-$ bandgap radiative relaxation with rate constant $R_z$) appear in solid arrows, while spin-mixing intraband spin relaxation (with timescales $\tau_{c,v,\ell}$ for conduction, valence and lower-valence bands, respectively, indicated by subscripts) appears in dashed arrows. The phonon-assisted $\Gamma_3^-\rightarrow\Gamma_1^+$ inter-valence-band transition (with time constant $\tau_h$) appears as a solid horizontal arrow.

We can write the following rate equations for the densities of conduction-band electrons $n_c^{\uparrow/\downarrow}$
\begin{align}
\frac{dn_c^\uparrow}{dt}=&X^\uparrow-\frac{n_c^\uparrow-n_c^\downarrow}{\tau_c}- R_z n_c^\uparrow p_v^\uparrow- R_\pm n_c^\uparrow p_\ell^\uparrow,\label{eq:rateeqns_first}\\ 
\frac{dn_c^\downarrow}{dt}=&X^\downarrow-\frac{n_c^\downarrow-n_c^\uparrow}{\tau_c}- R_z n_c^\downarrow p_v^\downarrow-R_\pm n_c^\downarrow p_\ell^\downarrow, 
\end{align}
and (lower) valence band holes  ($p_\ell^{\uparrow/\downarrow}$)  $p_v^{\uparrow/\downarrow}$:
\begin{align}
\frac{dp_v^\uparrow}{dt}=&\frac{p_\ell^\uparrow+p_\ell^\downarrow}{2\tau_h}- R_z n_c^\uparrow p_v^\uparrow-\frac{p_v^\uparrow-p_v^\downarrow}{\tau_v},\\ 
\frac{dp_v^\downarrow}{dt}=&\frac{p_\ell^\uparrow+p_\ell^\downarrow}{2\tau_h} -R_z n_c^\downarrow p_v^\downarrow -\frac{p_v^\downarrow-p_v^\uparrow}{\tau_v},\\ 
\frac{dp_\ell^\uparrow}{dt}=&X^\uparrow -\frac{p_\ell^\uparrow-p_\ell^\downarrow}{\tau_\ell}-\frac{p_\ell^\uparrow}{\tau_h}-R_\pm n_c^\uparrow p_\ell^\uparrow,\\ 
\frac{dp_\ell^\downarrow}{dt}=&X^\downarrow -\frac{p_\ell^\downarrow-p_\ell^\uparrow}{\tau_\ell}- \frac{p_\ell^\downarrow}{\tau_h}-R_\pm n_c^\downarrow p_\ell^\downarrow,
\end{align}
where the generation terms $X^{\uparrow/\downarrow}=G^{\uparrow/\downarrow}(D_\ell-p_\ell^{\uparrow/\downarrow})(D_c-n_c^{\uparrow/\downarrow})$ incorporate transition dependence on occupancy within a finite density of states $D_c$ and $D_\ell$. These  nonlinear equations are solved under steady-state conditions (all $\frac{d}{dt}=0$)  with the added constraint of global neutrality dictated by conservation of charge,
\begin{equation}
(n_c^\uparrow+n_c^\downarrow)-(p_v^\uparrow+p_v^\downarrow+p_\ell^\uparrow+p_\ell^\downarrow)=0.\label{eq:rateeqns_last}
\end{equation}

Example results for steady-state carrier densities are shown in Fig. \ref{fig:holepol}(b) as functions of excitation intensity $\propto G^\uparrow$, where rounded relative parameter values are chosen for illustrative purposes, and perfect optical orientation ($G^\downarrow=0$) is assumed. 
Because direct (electromagnetic radiative) transitions are faster than those that involve emission or absorption of large -momentum phonons during hole energy relaxation from $\Gamma_3^-$ to $\Gamma_1^+$, $R_zD_c$ is chosen to be more than an order of magnitude larger than the inter-valence band momentum relaxation rate $1/\tau_h$. Furthermore, as discussed in Sec. \ref{sec:soizero}, our tight-binding calculations show that the $\Gamma_2^-$ conduction band wavefunction has only 0.1\% minority spin-mixing probability compared to the $\Gamma_1^+$ valence band at 8\%, which is reflected in the relative values for the spin lifetimes $\tau_{c,v,l}$. 

It can be seen that at low excitation rates, the conduction and upper valence band carrier concentration are proportional to $\sqrt{G^\uparrow}$, whereas the lower valence band has a linear dependence. The former behavior is due to mutual Pauli exclusion suppressing recombination from these bands, whereas hole transitions from lower to upper valence band are controlled only by the initial carrier density.

With increasing excitation intensity, the carrier densities saturate. In this regime, the conduction band spin polarization $P_c$ approaches 100\%, and the upper valence band spin polarization $P_v$ deviates substantially from zero, as can be seen in Fig. \ref{fig:holepol}(c). Because $n_c^\uparrow \rightarrow D_c$ as $G^\uparrow \rightarrow \infty$ when $G^\downarrow =0$, an exact analytic expression for the saturation polarization can be obtained by solution to a polynomial equation of only quadratic order, but its utility to provide physical insight is limited by (grotesque) complexity. Nevertheless, under the assumption that $\tau_c\gg \tau_{v,h}$ and $D_cR_z\tau_v\gg 1$, we can simplify it to lowest order
\begin{equation}
P_v\approx -1+\frac{4}{D_cR_z\tau_v}.
\label{eq:approxpol}
\end{equation}
Using parameters from the simulation shown in Fig.~\ref{fig:holepol}, Eq.~(\ref{eq:approxpol}) yields upper valence band polarization of -92\%, in excellent agreement to the calculated $\approx -92.45$\%. Under the same assumptions, the lower valence band polarization $P_\ell\approx\frac{\tau_\ell}{2\tau_h+\tau_\ell}$ gives $+\frac{1}{3}$, again closely matching our numerical result of $\approx$+33.83\%.

\section{\label{sec:sum} Summary}

Our study of 2D $MX$ monolayers has revealed the fundamental origins behind several properties of its electronic structure. Throughout this group-theory analysis begun in Sec. \ref{sec:group}, symmetry-allowed momentum matrix elements have played an essential role in determining e.g. the $\mathbf{k\cdot\hat{p}}$ coupling leading to valence band `caldera', spin-orbit effect on band structure, orbital contribution to g-factor, optical absorption selection rules, and degree of carrier spin polarization from optical orientation with circularly-polarized electromagnetic radiation. We comment on the significance of each below.

As we saw in Sec. \ref{sec:mexhat}, the valence band caldera is formed through a competition between repulsive interactions from bands at higher and lower energies. Importantly, we found in Sec. \ref{sec:NFE} that since the eigenstates are confined to 2D, subbands originating from the primordial NFE [00] planewave can produce the $\Gamma_1^+$ highest valence band, {\em above} the [01]-rooted $\Gamma_3^\pm$ bands. The weak interband dipole interaction with these lower bands, the vanishing effect of the $\Gamma_2^-$ conduction band just above, and interactions from even higher bands ultimately results in the flat caldera. Two-dimensional confinement is the essential ingredient here, so this phenomenon is not expected to be unique to $MX$. In fact, although its symmetry is quite different from $MX$, phosphorene has an ultra-flat valence band along the zigzag direction for similar reasons.\cite{Li_PRB14}

Spin-orbit coupling induces several important features in the electronic structure. First of all, as detailed in Sec. \ref{sec:Dressel}, broken lattice inversion symmetry allows Dresselhaus spin splitting, and the remaining point-group symmetry excludes any in-plane effective field components and splitting dependence linear in $k$. More importantly, as discussed in Sec. \ref{sec:soizero}, the most obvious consequences of spin-orbit coupling are the broken twofold orbital degeneracy of the $\Gamma_3^\pm$ bands, and spin mixing into the band-edge states. The latter effect determines the scattering-induced spin relaxation rate, and the former enables optical orientation and an orbital contribution to the effective g-factor.  

The inequivalence of g-factor for states in the Brillouin zone toward $K$ and $K'$, discussed in Sec. \ref{sec:gfactor}, has the same root as valley-spin coupling in TMDCs. There, however, the valence band maxima occur not at the $\Gamma$-point, but at the $K$ and $K'$ points per se. Although we do indeed see a weak valley-dependent g-factor in $MX$, the six shallow valence band maxima on the `caldera' rim are invariably strongly coupled through intervalley scattering, making experimental confirmation through spin Hall effect or valley Zeeman effect especially challenging.

Because they determine the electromagnetic dipole symmetry, momentum matrix elements dictate the optical selection rules. We saw in Sec. \ref{sec:opt_ori} that in-plane momentum matrix elements connecting the $\Gamma_1^+$ and $\Gamma_2^-$ band-edge states are inherently weak and only due to spin-orbit-induced band mixing of $\Gamma_3^-$ character into the upper valence band.  Absorption of normally-incident transverse-polarized electromagnetic plane waves with $\hbar\omega$ at the bandgap energy is thus suppressed. This constraint on optical absorption does not exist in the three-dimensional bulk counterpart material, so that thinning to monolayer drastically attenuates optical absorption. It is interesting to contrast this situation to TMDCs, whose transition from indirect-gap in bulk to direct-gap in monolayer results in a massive {\em increase} in optical absorption.\cite{Mak_PRL10}

These optical selection rules can be exploited to generate spin-polarized electrons in the conduction band with circularly-polarized photons, as discussed in Sec. \ref{sec:absorb}. When carriers are excited via transitions from the $\Gamma_3^-$ lower valence band, relaxation down to the upper valence band (by emission of electromagnetic radiation polarized normal to the plane and propagating in-plane for electrons, and via non-spin-conserving cascade mediated by phonon emission for holes) results in oppositely-polarized $\Gamma_1^+$ upper-valence band holes, as discussed in Sec. \ref{sec:spinrelax}. 

We note that our application of group theory to analyze the symmetry properties of the $MX$ monolayer can be straightforwardly extended to the case of stacked few-layer 2D flakes. Depending on the stacking configuration and number of layers, the inversion asymmetry of the monolayer (endowing many of the important spin-dependent qualities discussed in this paper) may be retained. Otherwise, time-reversal symmetry will preserve spin degeneracy and the Dresselhaus splitting will disappear. 

Other potential extensions of our theoretical framework include an incorporation of the mechanical lattice dynamics to determine phonon symmetries that could be used to expose the qualitative and quantitative aspects of various momentum and spin relaxation mechanisms relevant to our work. For example, the definite parity of all IRs with respect to reflection symmetry $\sigma_h$ decouples flexural phonons (odd parity) from participating in spin-conserving momentum scattering, whereas their effect on EY spin-flip may be strong\cite{Song_PRL13}. The incorporation of extrinsic invariants into our theory, such as magnetic field, electric field, and strain, is straightforward once their transformation properties and corresponding IRs are identified. In particular, these external fields may break the in-plane mirror reflection symmetry of the monolayer lattice and induce linear Bychkov-Rashba spin-orbit coupling\cite{Bychkov_JETPL84}. The resulting interplay with the cubic Dresselhaus field may provide a potential scheme for electronically-actuated spin manipulation.  

\begin{acknowledgments}
We thank Dr. Yang Song for a careful reading of this manuscript, and Profs. H. Dery and I. \v{Z}uti\'c for initially encouraging us to undertake this study. Additionally, we thank Prof. E. Ott for valuable insight into analytic solution of the rate equations, and Prof. J.D. Sau for helpful comments on the g-factor calculation. This work was supported by the Office of Naval Research under contract N000141410317, the National Science Foundation under contract ECCS-1231855, and the Defense Threat Reduction Agency under contract HDTRA1-13-1-0013. Investigation into g-factor was enabled by NSF contract DMR-1408796.
\end{acknowledgments}

%

\end{document}